\documentclass[11pt]{article}
\newfont{\gothic}{eufm10 scaled\magstep 1}

\def\rp{\stackrel{\rightarrow}{\partial}}

\def\bigstar{\;\bigcirc\kern-0.9em\star\;}  
\begin{document}
\begin{flushright} 
{\sf hep-th/9810164 }\\ 
Miami TH/1/98 \\ 
ANL-HEP-PR-98-132
\end{flushright}

{\Large
\centerline{WIGNER TRAJECTORY CHARACTERISTICS  IN PHASE SPACE }
\centerline{AND FIELD THEORY}
\centerline{Thomas Curtright$^{\S}$ and Cosmas Zachos$^{\sharp}$} }

$^{\S}$ Department of Physics, University of Miami,
Box 248046, Coral Gables, Florida 33124, USA \\
\phantom{.} \qquad\qquad{\sl curtright@phyvax.ir.Miami.edu}  

$^{\sharp}$ High Energy Physics Division,
Argonne National Laboratory, Argonne, IL 60439-4815, USA \\
\phantom{.} \qquad\qquad{\sl zachos@hep.anl.gov}      
\begin{abstract}
Exact characteristic trajectories are specified for the time-propagating 
Wigner phase-space 
distribution function. They are especially simple---indeed, classical---for
the quantized simple harmonic oscillator, which serves as the underpinning 
of the field theoretic Wigner functional formulation introduced. Scalar field 
theory is thus reformulated in terms of distributions in field
phase space. Applications to duality transformations in field theory
are discussed.
\end {abstract}

\noindent\rule{7in}{0.02in}

An autonomous formulation of Quantum Mechanics, different from conventional 
Hilbert space or path integral quantization, is based on Wigner's 
phase-space distribution function  (WF), which is a special 
representation of the density matrix \cite{wigner}. In this formulation, known 
as deformation quantization \cite{bayen}, phase-space c-number functions are 
multiplied through the crucial  non-commutative $\star$-product \cite{groen}. 
The empowering principle underlying 
this quantization is its operational isomorphism \cite{bayen} to the 
conventional Heisenberg operator algebra of quantum mechanics.

Here, we employ the $\star$-unitary evolution operator, a 
``$\star$-exponential", to specify the time propagation of Wigner phase space 
distribution functions. The answer is known to be remarkably simple for the 
harmonic oscillator WF, and consists of classical rotation in phase-space 
for the 
full quantum system. It thus serves as the underpinning of the generalization 
to field theory we consider, in which the dynamics is specified through 
the evolution of c-number distributions on field phase space. 

Wigner functions are defined by 
\begin{equation}  
f(x,p)={\frac{1}{2\pi }}\int \!dy~\psi^{*}(x-{\frac{\hbar }{2}} 
y)~e^{-iyp}\psi(x+{\frac{\hbar }{2}}y)\;.
\end{equation} 
Even though they amount to spatial auto-correlation functions of 
Schr\"{o}dinger wavefunctions $\psi$, they can be determined without reference 
to such wavefunctions, in a logically autonomous structure. For instance, 
when the wavefunction is an energy ($E$) eigenfunction,
the corresponding WF is time-independent and satisfies the two-sided 
energy $\star $-genvalue equations \cite{dbfcam,cfz}, 
\begin{equation}  
H\star f=f\star H=E f \;,
\end{equation} 
where $\star $ is the essentially unique associative deformation of 
ordinary products on phase-space,
\begin{equation}
\star \equiv e^{{\frac{i\hbar }{2}}(\stackrel{\leftarrow }{\partial }_{x}%
\stackrel{\rightarrow }{\partial }_{p}-\stackrel{\leftarrow }{\partial }_{p}%
\stackrel{\rightarrow }{\partial }_{x})}\;,
\end{equation}
as defined by Groenewold \cite{groen}, and developed in \cite{bayen}.
In practice, it may be evaluated through translations of function arguments, 
$f(x,p) \star g(x,p) = f(x+{i\hbar\over 2}\rp_p ,~ p-{i\hbar\over 2}\rp_x)~ 
g(x,p)$, to produce pseudodifferential equations. 

These WFs are real. They are bounded by the Schwarz inequality \cite{baker} to 
$-2/h\leq f\leq2/h$. They can go negative, and, indeed, they do for all but 
Gaussian configurations, so they are not probability distributions 
\cite{wigner}. However, upon integration over $x$ or $p$, they  yield
marginal probability densities in $p$ and $x$-space, respectively.
They can also be shown to be orthonormal \cite{dbfcam,cfz}. 
Unlike in Hilbert space quantum mechanics, naive superposition of solutions 
of the above does not hold, because of Baker's \cite{baker} fundamental 
nonlinear projection condition $f\star f=f/h$. 

Time-dependence for WFs was succintly specified by Moyal through the commutator 
bracket \cite{moyal} bearing his name, 
\begin{equation} 
i\hbar \frac{\partial }{\partial t}\,f(x,p;t)=H\star f(x,p;t)-f(x,p;t)\star H.
\label{jomo}
\end{equation}
This turns out to be the essentially unique associative generalization of the 
Poisson  Bracket \cite{vey}, to which it reduces as $\hbar\rightarrow 0$, 
yielding Liouville's 
theorem of classical mechanics, 
${\partial}_t \,f + \{ f,H \} =0$.  

For the evolution of the fundamental phase-space 
variables $x$ and $p$, time evolution is simply the convective part 
of Moyal's equation, so the apparent sign is reversed, while the Moyal Bracket 
actually reduces to the Poisson Bracket.
That is, the $\hbar$-dependence
drops out, and these variables, in fact, evolve simply by the 
{\em classical} Hamilton's equations of motion, $\dot{x}= \partial_p H$,
$\dot{p}= -\partial_x H$. 

What is the time-evolution of a WF like? This is the first question 
we address.
Relying on the isomorphism to operator algebras of \cite{bayen} 
indicated, one may solve for the time-trajectories of 
the WF, which turn out to be notably simple.
By virtue of the $\star$-unitary evolution operator, 
a ``$\star$-exponential" \cite{bayen},
\begin{equation} 
U_{\star} (x,p;t)=e_\star ^{itH/\hbar} \equiv 
1+(it/\hbar)H(x,p) + {(it/\hbar)^2\over 2!}  H\star H +{(it/\hbar )^3\over 3!} 
 H\star H\star H +...,
\end{equation}
the time-evolved Wigner function is obtainable formally in 
terms of the Wigner function at $t=0$ through associative combinatoric
operations completely analogous to the conventional formulation of quantum 
mechanics of operators in Hilbert space.
Specifically, 
\begin{equation} 
f(x,p;t)=U_{\star}^{-1} (x,p;t) \star f(x,p;0)\star U_{\star}(x,p;t) .    
\label{evol}
\end{equation}
As mentioned, the dynamical variables evolve classically, 
\begin{equation}  
{dx \over dt}= {x \star H-H\star x \over  i\hbar}= \partial_p H ~,
\end{equation} 
and 
\begin{equation}  
{dp \over dt}= {p \star H-H\star p \over i\hbar}=- \partial_x H ~.
\end{equation} 
Consequently, by associativity, the quantum evolution,
\begin{equation}  
x(t)=U_{\star} \star x \star U_{\star}^{-1}, 
\end{equation} 
\begin{equation}  
p(t)=U_{\star} \star p \star U_{\star}^{-1}, 
\end{equation} 
turns out to flow along {\em classical} trajectories. 

What can one say about this formal time-evolution expression?
Any WF in phase space, upon Fourier transformation resolves to 
\begin{equation} 
f(x,p)= \int\! da db ~\tilde{f}(a,b) ~ e^{iax} e^{ibp}.
 \end{equation} 
However, note that  exponentials of individual functions of 
$x$ and $p$ are also $\star$-exponentials of the same functions, or 
 $\star$-versions of these functions, since the $\star$-product trivializes
in the absence of a conjugate variable, so that 
\begin{equation}
e^{iax} ~e^{ibp}=e_{\star} ^{iax} ~e_{\star} ^{ibp}  . \label{grolp}
\end{equation} 
Moreover, this is proportional to a $\star$-product, since  
\begin{equation}  
e_{\star} ^{iax}\star  e_{\star} ^{ibp}=
e_{\star} ^{ia(x+i\hbar \rp_p/2)} ~ e_{\star} ^{ibp}
=e_{\star} ^{iax} e_{\star} ^{ibp} e^{-i\hbar ab / 2}. \label{grulp}
\end{equation} 

Consequently, any Wigner function can be written as 
\begin{equation} 
f(x,p)= \int\! da db ~\tilde{f}(a,b) ~e^{i\hbar ab/ 2} ~
e_{\star} ^{iax}\star  e_{\star} ^{ibp}.
\end{equation} 
It follows then, that, by insertion of $U_{\star}\star U_{\star}^{-1}  $ 
pairs at every $\star$-multiplication, in general, 
$$
f(x,p;t)=\int\! da db ~\tilde{f}(a,b) ~e^{{i\hbar ab /2}} ~
e_{\star} ^{iaU_{\star}^{-1}   \star x \star U_{\star}}\star   ~  
e_{\star} ^{ibU_{\star}^{-1}  \star p \star U_{\star}} 
$$
\begin{equation} 
=\int\! da db ~\tilde{f}(a,b) ~e^{i\hbar ab /2}~
e_{\star} ^{iax(-t)}\star  e_{\star} ^{ibp(-t)}.
\end{equation} 

Unfortunately, in general, the above steps cannot be simply reversed to yield 
an integrand of the  form $\tilde{f}(a,b) ~ e^{iax(-t)} e^{ibp(-t)}$. 
But, in some fortuitous circumstances, they can, and in this case the evolution
of the Wigner function reduces to merely backward evolution of its arguments 
$x,p$ along classical trajectories, while its functional form itself remains 
unchanged:
\begin{equation} 
f(x,p;t)=f\left( x(-t),p(-t);0\right).  \label{gold}
 \end{equation} 

To illustrate this, consider the simple linear harmonic oscillator 
(taking $m=1$, $\omega=1$),
\begin{equation}  
H={p^2 +x^2\over 2}= \frac{x-ip} {\sqrt{2}} \star \frac{x+ip} {\sqrt{2}} +
{\hbar\over 2} ~.\label{simplex}
\end{equation} 
It is easily seen that 
\begin{equation}  
i\hbar \dot{x}= x \star H-H\star x = i\hbar p ~ , \qquad 
i\hbar \dot{p}= p \star H-H\star p = -i\hbar x   ~,
\end{equation} 
and thus the canonical variables indeed evolve classically: 
\begin{equation}  
X\equiv x(t)=U_{\star}\star x\star U_{\star}^{-1}= x\cos t + p \sin t, \qquad  
P\equiv p(t)=U_{\star}\star p\star U_{\star}^{-1}= p\cos t - x \sin t.  
\label{rotation} 
\end{equation} 
This checks against the $\star$-exponential for the SHO, \cite{bayen}, \qquad
$e_\star ^{itH/\hbar}=\frac{1} {\cos (t/2)} \exp ( {2i\tan (t/2)} H/{\hbar} )$.

Now, recall the degenerate case of the Baker-Campbell-Hausdorff 
combinatoric identity for any two operators with {\em constant} commutator 
with respect to any associative multiplication, thus for
any phase-space functions $\xi$ and $\eta$ under $\star$-multiplication. If
\begin{equation} 
\xi \star \eta - \eta \star \xi=c,
 \end{equation} 
then, 
\begin{equation} 
e_{\star} ^{\xi}\star  e_{\star} ^{\eta}=
e_{\star} ^{\xi+\eta} ~e^{c/2}.
\end{equation} 
Application of this identity as well as (\ref{grulp}) and (\ref{grolp}) 
yields directly
\begin{eqnarray} 
e_{\star}^{iax(-t)} \star e_{\star} ^{ibp(-t)}   ~e^{i\hbar ab/2}&  =&
e_{\star}^{i(a\cos t+b\sin t ) x+ i(b\cos t-a\sin t)p }\nonumber \\
&=& e_{\star}^{i(a\cos t+b\sin t ) x } \star e_{\star} ^{i(b\cos t-a\sin t)p } 
~e^{i\hbar (a\cos t+b\sin t) (b\cos t-a \sin t)/2}  \nonumber \\
&=& e_{\star}^{i(a\cos t+b\sin t ) x } e_{\star} ^{i(b\cos t-a\sin t)p}
\nonumber \\
&=& e^{i(a\cos t+b\sin t ) x }~e^{i(b\cos t-a\sin t)p }~.  
 \end{eqnarray} 
Consequently,
\begin{equation} 
f(x,p;t)= \int\! da db ~\tilde{f}(a,b) ~ e^{iax(-t)} e^{ibp(-t)},
 \end{equation} 
and hence the reverse convective flow (\ref{gold}) obtains.

The result for the SHO is the preservation of the functional form of the Wigner 
distribution function along classical phase-space trajectories:
\begin{equation}  
f(x,p;t)=f(x \cos t - p \sin t,     p \cos t + x \sin t  ;0). \label{preserv}
\end{equation} 
What this means is that {\em any} Wigner distribution rotates 
uniformly on the phase plane around the origin, essentially classically,
even though it provides a complete quantum mechanical description. Note how, 
in general,  this result is deprived of import, or, at the very least, 
simplicity, upon integration in $x$ (or $p$)
to yield probability densities: the rotation induces shape variations of the 
oscillating probability density profile. Only if, as is the case for 
coherent states \cite{almeida}, a Wigner 
function configuration has an additional 
axial $x-p$ symmetry around its {\em own} center, will it possess an invariant 
profile upon this rotation, and hence a shape-invariant oscillating probability 
density.

The result (\ref{preserv}), of course, is not new. It was clearly recognized 
by Wigner \cite{kim}. 
It follows directly from (\ref{jomo}) for (\ref{simplex}) that 
\begin{equation}  
\left(  \partial_t +p \partial_x- x \partial_p\right) f(x,p;t)=0~. 
\label{character} 
\end{equation} 
The characteristics of this partial differential equation correspond to the 
above uniform rotation in phase space, so it is easily seen to be solved 
by (\ref{preserv}).   The result was given 
explicitly in \cite{groen} and also \cite{bartlett}, following
different derivations. B Lesche \cite{lesche},
has also reached this result in an elegant fifth derivation, 
by noting that for quadratic Hamiltonians such as this one,
the linear rotation of the dynamical variables (\ref{rotation}) 
leaves the symplectic quadratic form invariant, and thus the $\star$-product 
invariant. That is, the gradients in the $\star$-product may also be taken to 
be with respect to the time-evolved canonical variables (\ref{rotation}),
$X$ and $P$; 
hence, after inserting $U_{\star}\star U_{\star}^{-1}  $ in the $\star$-
functional form of $f$, the $\star$-products may be taken to be with respect to 
$X$ and $P$, and the functional form of $f$ is preserved, 
(\ref{gold}). This only holds for quadratic Hamiltonians. 

Dirac's interaction representation may then be based on this property,
for a general Hamiltonian consisting of a basic SHO part, 
$H_0=(p^2 +x^2)/2$, supplemented by an interaction part,
\begin{equation}
H= H_0 + H_I.
\end{equation} 
Now, upon defining
\begin{equation}
w\equiv e_\star ^{itH_0/\hbar} \star f \star e_\star ^{-itH_0/\hbar} ,
\end{equation} 
it follows that Moyal's evolution equation reads,
\begin{equation}
i\hbar \frac{\partial }{\partial t}\,w(x,p;t)
={\cal H}_I \star w (x,p;t)- w(x,p;t)\star {\cal H}_I,
  \end{equation} 
where 
${\cal H}_I \equiv e_\star ^{itH_0/\hbar}\star H_I \star 
e_\star ^{-itH_0/\hbar}$. Expressing $H_I$ as a $\star$-function leads to 
a simplification. 

In terms of the convective variables (\ref{rotation}), $X,P$,
${\cal H}_I(x,p) =H_I (X,P)$, and $w(x,p;t) =f(X,P;t)$, while 
$\star$ may refer to these convective variables as well. Finally, then,
\begin{equation}
i\hbar \frac{\partial }{\partial t}\,f(X,P;t)
= H_I(X, P) \star f (X, P;t)- f(X, P;t)\star H_I (X, P).  \label{interaction}
\end{equation} 
In the uniformly rotating frame of the convective variables, the WF 
time-evolves according to the interaction Hamiltonian---while, for vanishing 
interaction Hamiltonian,  $f (X, P;t)$ is constant and yields 
(\ref{preserv}). Below, in generalizing to field theory, 
this provides the basis of the interaction picture of perturbation theory,
where the basis canonical fields evolve classically as above \footnote{
Ref \cite{nachbag} discusses a field theoretic interaction representation
in phase space, which does not appear coincident with the present one.}.
 
To produce Wigner functionals in scalar field theory, one may start from the 
standard, noncovariant, formulation of field theory in Hilbert space, 
in terms of Schr\"{o}dinger wave-functionals. 

For a free field Hamiltonian, the energy eigen-functionals are Gaussian in 
form. For instance, without loss of generality, in two dimensions ($x$ is a 
spatial coordinate, and $t=0$ in all fields),
the ground state functional is 
\begin{equation}
\Psi[ \phi] =\exp\left(  -\frac{1}{2\hbar}\int dx\,\phi\left(
x\right)  \sqrt{m^{2}-\nabla_{x}^{2}}~ \phi\left(  x\right)  \right).
\end{equation}
Boundary conditions are assumed such that the $\sqrt{m^{2}-\nabla_{x}^{2}}$
kernel in the exponent is naively self-adjoint. 
``Integrating by parts'' one of the $\sqrt{m^{2}-\nabla_{z}^{2}}$ kernels,
functional derivation $\delta\phi\left(  x\right)  /\delta\phi\left(  z\right)
=\delta\left(  z-x\right) $ then leads to  
\begin{equation}  
\hbar\frac{\delta}{\delta\phi\left(  z\right)  }\Psi [\phi] 
=-\left(  \sqrt{m^{2}-\nabla_{z}^{2}}\,\phi\left(  z\right)  \right)
\Psi[ \phi] ,
\end{equation} 
\begin{equation}  
\hbar^{2}\frac{\delta^{2}}{\delta\phi\left(  w\right)  \delta\phi\left(
z\right)  }\Psi[ \phi] =\left(  \sqrt{m^{2}-\nabla_{w}^{2}}\,
\phi\left(  w\right)  \right)  \left(  \sqrt{m^{2}-\nabla_{z}^{2}}\,
\phi\left(  z\right)  \right)  \,\Psi[ \phi] -\hbar\sqrt
{m^{2}-\nabla_{z}^{2}}\,\delta\left(  w-z\right)  \,\Psi[ \phi]. 
\end{equation} 

Note that the divergent zero-point energy density, 
\begin{equation}  
E_0= {\hbar\over 2  }    \lim_{w\rightarrow z} 
\sqrt {m^{2}-\nabla_{z}^{2}}~ \delta\left(  w-z\right), 
\end{equation} 
may be handled rigorously using $\zeta$-function regularization. 

Leaving this zero-point energy present,  leads to the 
standard energy eigenvalue equation, again through integration by parts, 
\begin{equation}  
\frac{1}{2} \int dz\,\left(  -\hbar^{2}\frac{\delta^{2}}
{\delta\phi\left(  z\right)  ^{2} }+\phi\left(  z\right)  
\left(  m^{2}-\nabla_{z}^{2}\right)  \phi\left( z\right)  \right)  \Psi[ \phi ] 
=E_0 ~\Psi[ \phi]. 
\end{equation} 

A natural adaptation to the corresponding Wigner functional is the following.
For a functional measure $\left[ d\eta / 2\pi \right]  =\prod_{x} 
d\eta\left(  x\right)  / 2\pi$, one obtains
\begin{equation}
W [\phi,\pi] =\int\!\left[  \frac{d\eta}{2\pi}\right]
~\Psi^{\ast}\left[  \phi-{\frac{\hbar}{2}\eta}\right]  \,
e^{-i\int dx\,\eta\left(  x\right)  \pi\left(  x\right)  }~
\Psi\left[  \phi+{\frac{\hbar }{2}\eta}\right] ,
\end{equation}
where $\pi\left(  x\right)  $ is to be understood as the local field variable
canonically conjugate to $\phi\left(  x\right)  $. However, in this
expression, both $\phi$ and $\pi$ are {\em classical} variables, not
operator-valued fields, in full analogy to the phase-space quantum mechanics 
already discussed. 

For the Gaussian ground state wavefunctional, this evaluates to 
\begin{eqnarray}
W\left[  \phi,\pi\right]   & =&\int\left[  \frac{d\eta}{2\pi}\right]
\exp\left(  -\frac{1}{2\hbar}\int dx\,\left(  \phi\left(  x\right)
-{\frac{\hbar}{2}\eta}\left(  x\right)  \right)  \,\sqrt{m^{2}-\nabla_{x}^{2}%
}\,\left(  \phi\left(  x\right)  -{\frac{\hbar}{2}\eta}\left(  x\right)
\right)  \right)  ~\times\!\\
~& \times&~e^{-i\int dx\,\eta\left(  x\right)  \pi\left(  x\right)  }%
\exp\left(  -\frac{1}{2\hbar}\int dx\,\left(  \phi\left(  x\right)
+{\frac{\hbar}{2}\eta}\left(  x\right)  \right)  \,\sqrt{m^{2}-\nabla_{x}^{2}%
}\,\left(  \phi\left(  x\right)  +{\frac{\hbar}{2}\eta}\left(  x\right)
\right)  \right)   \nonumber 
\end{eqnarray}
$$ =\exp\left(  -\frac{1}{\hbar}\int dx\,\phi\left(  x\right)  \,\sqrt
{m^{2}-\nabla_{x}^{2}}\,\phi\left(  x\right)  \right)  
\left(  \int\left[  \frac{d\eta}{2\pi}\right]  \,e^{-i\int
dx\,\eta\left(  x\right)  \pi\left(  x\right)  }\exp\left(  -\frac{\hbar}%
{4}\int dx\,{\eta}\left(  x\right)  \,\sqrt{m^{2}-\nabla_{x}^{2}}\,{\eta
}\left(  x\right)  \right)  \right).
$$
So
\begin{equation}
W[\phi,\pi] =\mathcal{N}\exp\left(  -\frac{1}{\hbar}\int 
dx\,\left(  \left(  \phi\left(  x\right)  \,\sqrt{m^{2}-\nabla_{x}^{2}}\,
\phi\left(  x\right)\right) + \left(  {\pi}\left(  x\right)  \,\left(
\sqrt{m^{2}-\nabla_{x}^{2}}\right) ^{-1} {\pi}\left(  x\right)  \right) 
\right) \right) ,
\end{equation}
where $\mathcal{N}$ is a normalization factor. It is the expected collection of 
harmonic oscillators. 

This Wigner functional is, of course \cite{cfz}, an energy
$\star$-genfunctional, also checked directly. For
\begin{equation}  
H_0[\phi,\pi]\equiv \frac{1}{2} \int dx\left(  \pi\left(  x\right) ^{2}
+\phi\left(  
x\right)  \left( m^{2}-\nabla_{x}^{2}\right)  \phi\left(  x\right)  \right),
\end{equation} 
and the inevitable generalization 
\begin{equation}
\star \equiv \exp \left( {\frac{i\hbar }{2}}\int dx~ \left( 
\stackrel{\leftarrow } {\delta\over\delta\phi (x)} 
\stackrel{\rightarrow }{\delta\over\delta\pi (x)}
-\stackrel{\leftarrow }{\delta\over\delta\pi (x)}
\stackrel{\rightarrow }{\delta\over\delta\phi (x)}\right)  \right)  ~,
\end{equation}  
it follows that 
\begin{eqnarray}
H_0\star W  &=&\int \frac{dx}{2}     \left(
\left(  \pi\left(  x\right)  -\frac{1}{2}i\hbar\frac{\delta}{\delta\phi\left(
x\right)  }\right)  ^{2}
+\left(  \phi\left(  x\right)  +\frac{1}{2}i\hbar\frac{\delta}{\delta
\pi\left(  x\right)  }\right)  \left(  m^{2}-\nabla_{x}^{2}\right)  \left(
\phi\left(  x\right)  +\frac{1}{2}i\hbar\frac{\delta}{\delta\pi\left(
x\right)  }\right)\right)  W\left[  \phi,\pi\right]  \nonumber \\  
& =&\int\frac{dx }{2}  \left(
\pi\left(  x\right)  ^{2}-\frac{1}{4}\hbar^{2}\frac{\delta}{\delta\pi\left(
x\right)  }\left(  m^{2}-\nabla_{x}^{2}\right)  \frac{\delta}{\delta\pi\left(
x\right)  } 
+\phi\left(  x\right)  \left(  m^{2}-\nabla_{x}^{2}\right)  \phi\left(
x\right)  -\frac{1}{4}\hbar^{2}\frac{\delta^{2}}{\delta\phi\left(  x\right)
^{2}} \right)  W\left[  \phi,\pi\right]  \nonumber\\
&=& E_0 \,W\left[\phi ,\pi\right].
\end{eqnarray}
This is indeed the ground-state Wigner energy-$\star$-genfunctional. 
The $\star$-genvalue is again the zero-point energy, which could have been 
removed by point-splitting the energy density, as indicated earlier. 
There does not seem to be a simple point-splitting procedure that 
regularizes the star product as defined above and also preserves associativity.

As in the case of the SHO discussed above, free-field time-evolution for 
Wigner functionals is also effected by Dirac
delta functionals whose support lies on the classical field time evolution
equations. 
Fields evolve according to the equations,
\begin{equation}  
-i\hbar\partial_{t}\phi  =H\star\phi-\phi\star H,\qquad \qquad 
-i\hbar\partial_{t}\pi  =H\star\pi-\pi\star H.
\end{equation} 
For $H_0$, these equations are the classical evolution
equations for free fields, 
\begin{equation}  
\partial_{t}\phi\left(  x,t\right)     =\pi\left(  x,t\right) , \qquad \qquad 
\partial_{t}\pi\left(  x,t\right)  =-\left(  m^{2}-\nabla_{x}^{2}\right)
\phi\left(  x,t\right).
\end{equation} 

Formally, the solutions are represented as 
\begin{eqnarray}
\phi\left(  x,t\right)    & =&\cos\left(  t\sqrt{m^{2}-\nabla_{x}^{2}}\right)
\phi\left(  x,0\right)  +\sin\left(  t\sqrt{m^{2}-\nabla_{x}^{2}}\right)
\frac{1}{\sqrt{m^{2}-\nabla_{x}^{2}}}~ \pi\left(  x,0\right)  \\
\pi\left(  x,t\right)    & =&-\sin\left(  t\sqrt{m^{2}-\nabla_{x}^{2}}\right)
\sqrt{m^{2}-\nabla_{x}^{2}}~\phi\left(  x,0\right)  +\cos\left(  t\sqrt
{m^{2}-\nabla_{x}^{2}}\right)  \pi\left(  x,0\right).
\end{eqnarray}
From these, it follows by the functional chain rule that
$$
\int dx\left(  \pi\left(  x,0\right)  \frac{\delta}{\delta\phi\left(
x,0\right)  }-\left(  \left(  m^{2}-\nabla_{x}^{2}\right)  \phi\left(
x,0\right)  \right)  \,\frac{\delta}{\delta\pi\left(  x,0\right)  }\right)
\qquad \qquad \qquad \qquad 
$$
\begin{equation}  
\qquad \qquad \qquad 
\qquad =\int dx\left(  \pi\left(  x,t\right)  \frac{\delta}{\delta\phi\left(
x,t\right)  }-\left(  \left(  m^{2}-\nabla_{x}^{2}\right)  \phi\left(
x,t\right)  \right)  \,\frac{\delta}{\delta\pi\left(  x,t\right)  }\right)
\end{equation} 
for any time $t$. 

Consider the free-field Moyal evolution equation for a generic (not necessarily 
energy-$\star$-genfunctional) WF, corresponding to (\ref{character}), 
\begin{equation}
\partial_{t}W  = -\int dx\,\left(  \pi\left(  x\right)  \frac{\delta}{\delta
\phi\left(  x\right)  }-\phi\left(  x\right)  \left(  m^{2}-\nabla_{x}%
^{2}\right)  \frac{\delta}{\delta\pi\left(  x\right)  }\right)  W.
\end{equation}
The solution is
\begin{equation}  
W\left[  \phi,\pi;t\right]  =W\left[  \phi\left(  -t\right)  ,\pi\left(
-t\right)  ;0\right].
\end{equation} 

Adapting the method of characteristics for
first-order equations to a functional context,
one may simply check this solution again using the chain rule for functional 
derivatives, and the
field equations {\em evolved backwards in time} as specified:
$$
\partial_{t}W\left[  \phi,\pi;t\right]     =\partial_{t}W\left[  \phi\left(
-t\right)  ,\pi\left(  -t\right)  ;0\right]  
\qquad\qquad\qquad\qquad\qquad\qquad\qquad\qquad\qquad\qquad
$$ 
\begin{eqnarray}
& =&\int dx\,\left(  \partial
_{t}\phi\left(  x,-t\right)  \,\frac{\delta}{\delta\phi\left(  x,-t\right)
}+\partial_{t}\pi\left(  x,-t\right)  \,\frac{\delta}{\delta\pi\left(
x,-t\right)  }\right)  
W\left[  \phi\left(  -t\right)  ,\pi\left(  -t\right);0\right]  \nonumber \\
& =&\int dx\left(  \,\left(  -\pi\left(  x,-t\right)  \right)  \frac{\delta
}{\delta\phi\left(  x,-t\right)  }+\left(  \left(  m^{2}-\nabla_{x}%
^{2}\right)  \phi\left(  x,-t\right)  \right)  \,\frac{\delta}{\delta
\pi\left(  x,-t\right)  }\right)  W\left[  \phi\left(  -t\right)  ,\pi\left(
-t\right)  ;0\right]\nonumber   \\
& =&\int dx\left(  \,\left(  -\pi\left(  x,-t\right)  \right)  \frac{\delta
}{\delta\phi\left(  x,-t\right)  }+\left(  \left(  m^{2}-\nabla_{x}%
^{2}\right)  \phi\left(  x,-t\right)  \right)  \,\frac{\delta}{\delta
\pi\left(  x,-t\right)  }\right)  W\left[  \phi,\pi;t\right] \nonumber \\
& =&-\int dx\,\left(  \pi\left(  x\right)  \frac{\delta}{\delta\phi\left(
x\right)  }-\left(  m^{2}-\nabla_{x}^{2}\right)  \phi\left(  x\right)
\,\frac{\delta}{\delta\pi\left(  x\right)  }\right)  W\left[  \phi
,\pi;t\right ].
\end{eqnarray}
The quantum Wigner Functional for free fields time-evolves along classical  
field configurations. In complete analogy to the interaction 
representation for single particle quantum mechanics, (\ref{interaction}), the 
perturbative series in the interaction Hamiltonian (written as a 
$\star$-function of fields) is then defined
in terms of convective (time-evolved free field) variables $\Phi,\Pi$:
\begin{equation}
i\hbar \frac{\partial }{\partial t}\,W[\Phi,\Pi;t]
= H_I[\Phi,\Pi] \star W [\Phi,\Pi ;t]- W[\Phi,\Pi ;t]\star 
H_I [\Phi,\Pi]~.  
\end{equation} 

In ref \cite{cfz}, a transformation function $T$ was introduced to 
accommodate arbitrary canonical transformations induced by a generating 
function $F$ in quantum mechanics, following Dirac. 
The WF in terms of the canonically transformed variables is obtained by 
convolving with this transformation function. In complete analogy, in 
scalar field theory, cf \cite{pcm}, given a canonical transformation from field 
variables $\phi,\pi$ to variables $\varphi,\varpi$ effected by a 
generating functional $F[\phi, \varphi]$, one may 
deduce that the WF in terms of the canonically transformed field variables is 
\begin{equation}  
W[\phi,\pi]  =\int\!\left[  \frac{d\varphi d\varpi}{2\pi}\right]
T[\phi,\pi;\varphi,\varpi] ~ \mathcal{W}\left[\varphi,\varpi\right], 
\end{equation} 
where 
\begin{equation}
T[\phi,\pi;\varphi,\varpi]=\! \int\! \left[  \frac{d\eta d\rho}{2\pi}\right]
\exp i\left(
F[\phi+\frac{1}{2}\eta,\varphi+{\frac{1}{2}\rho}]-iF^{*}[\phi-{\frac
{1}{2}\eta},\varphi-{\frac{1}{2}\rho}]+\! \int\! dx (  \varpi\left(  x\right)
  \rho\left(  x\right)  -\pi\left( x\right)  \eta\left(  x\right) ) \right).
\end{equation}

For example, the generating functional for free field duality between 
a two-dimensional
spacetime scalar $\varphi$ and a pseudoscalar $\phi$ is
\begin{equation}
F[\phi,\varphi]=\int\! dx~ \phi\,\partial_{x}\varphi,
\end{equation}
so it yields the classical canonical transformations 
\begin{equation}  
\pi  =\frac{\delta}{\delta\phi}F=\partial_{x}\varphi, \qquad \qquad 
\varpi  =-\frac{\delta}{\delta\varphi}F =\partial_{x}\phi.
\end{equation} 
After some computation, it follows that 
\begin{equation}  
T[\phi,\pi;\varphi,\varpi]  = \left[  2\pi\right]  \,\delta\left[  
\partial_{x}\varphi-\pi\right] ~
\delta\left[  \varpi-\partial_{x}\phi\right].
\end{equation} 
The ensuing relation between the respective dual Wigner functionals is then
quite simple:
\begin{equation}  
W[\phi,\pi]   =\mathcal{W}\left[\int^{x}\! \pi,~\partial_{x}\phi\right].
\end{equation} 

A less exceptional example is the canonical transformation from the 
chiral $\sigma$-model in two dimensions to its dual counterpart, \cite{pcm}, 
generated by 
\begin{equation}  
F=\int\! dx~\phi^{i}J^{i}\left(  \varphi\right),
\end{equation} 
where
\begin{equation}  
J^{i}\left[  \varphi\right]  =\sqrt{1-\varphi^{2}}~ 
\stackrel{\leftrightarrow}{\partial_x} 
\varphi^{i}+\varepsilon^{ijk}\varphi^{j}\partial_{x}\varphi^{k}%
\end{equation} 
is the spatial component of the right (V+A) current. The resulting 
transformation functional is more involved than above,
$$
T = \int\!\left[  d\rho\right]  \,\delta\left[  \pi^{i}-\frac{1}{2}J^{i}\left(
\varphi+{\frac{1}{2}\rho}\right)  -\frac{1}{2}J^{i}\left(  \varphi-{\frac
{1}{2}\rho}\right)  \right]  \times  \qquad\qquad \qquad\qquad \qquad\qquad 
$$
\begin{equation}  
\qquad\qquad \qquad\qquad \qquad\qquad \times 
\exp\left(  i\int dx\phi^{i}\left(  J^{i}\left(  \varphi+{\frac{1}%
{2}\rho}\right)  -J^{i}\left(  \varphi-{\frac{1}{2}\rho}\right)  \right)
\right)  \exp\left(i\int dx\varpi \rho \right).
\end{equation} 

\noindent{\Large{\bf Acknowledgements}} 

This work was supported in part by NSF grant PHY 9507829, and
by the US Department of Energy, Division of High Energy Physics, 
Contract W-31-109-ENG-38. TLC thanks the Fermilab and Argonne theory groups 
for their hospitality and summer support.

\noindent\rule{7in}{0.02in}

\end{document}